\documentclass[prb,preprint]{revtex4}

\usepackage{graphicx}
\usepackage{dcolumn}
\usepackage{bm}

\begin{document}

\title{
Dynamical Brittle Fractures of Nanocrystalline Silicon \\
using Large-Scale Electronic Structure Calculations \\
}
\author{Takeo Hoshi}
\author{Takeo Fujiwara}

\affiliation{Department of Applied Physics, University of Tokyo,   Bunkyo-ku, Tokyo 113-8656, Japan}   

\begin{abstract}
A hybrid scheme between
large-scale electronic structure calculations
is developed and applied to 
nanocrystalline silicon with more than 10$^5$ atoms.
Dynamical fracture processes 
are simulated under external loads in the [001] direction.
We shows that the fracture propagates anisotropically on 
the (001) plane and reconstructed surfaces 
appear with asymmetric dimers.
Step structures are formed in larger systems,
which is understood as
the beginning of a crossover 
between nanoscale and macroscale samples.
\end{abstract} 

\sloppy
\maketitle


Silicon is one of ideally brittle materials 
and its fracture behavior is studied intensively, 
because we can obtain
essentially dislocation-free single crystals.
A pioneering theory 
of brittle fracture was given, in 1920's,
by Griffith \cite{GRIFFITH} within a continuum theory, 
which is the foundation of the present understanding of 
brittle fractures \cite{BRITTLE-TEXTBOOK}.
The fracture in single crystals
should be investigated also in atomistic pictures,  
on the point of how and why the fracture path 
is formed and propagates in the crystalline geometry.
This point includes surface reconstruction processes. 
Since fracture is a thermal non-equilibrium process, 
the atomic structure on a cleavage surface 
can be different from that on equilibrium clean surfaces.
For example, 
the easiest cleavage plane in macroscale samples of silicon 
is the (111) plane, in which
the surface structure is 
not the ground state ($7 \times 7$) structure
but a metastable $2 \times 1$ structure
\cite{PANDEY1,SI111-21-PARRINELLO-MD}.

This letter is devoted to 
the atomistic fracture behaviors
in {\it nanocrystalline} silicon,
especially,
its possible difference from macroscale samples.
Such a difference can be expected, as explained below;
Now a typical atomistic length in silicon is defined as
$d_0 \equiv \, ^3 \! \! \! \sqrt{v_0} \approx 3$ \AA, 
where $v_0$ gives the volume per atom.
The essence of the Griffith theory \cite{GRIFFITH} is
the energy competition
between the energy gain of the strain relaxation 
and the loss of the surface formation energy.
The former energy 
is a volume term proportional to (length)$^3$,
while the latter energy
is a surface term proportional to (length)$^2$.
As analogous 
to the theory of nucleation \cite{LANDAU},
the dimensional analysis gives
the critical crack length
for the spontaneous fracture propagation. 
The critical crack length  $c_{\rm G}$ is given as
\cite{GRIFFITH,BRITTLE-TEXTBOOK}
\begin{eqnarray}
 c_G \approx \frac{\gamma E}{\sigma^2}
 \label{GRIFFITH-CG}
\end{eqnarray}
with the stress $\sigma$,
the Young modulus $E(\approx 10^2{\rm GPa})$ and 
the surface energy per unit area $\gamma$.
The value of $\gamma$ was estimated to be in the order of
$1 {\rm J/m^2}$ \cite{SPENCE93,GUMBSCH2},
which  can be reduced to the bond breaking energy
$(\gamma d_0^2 \approx 1{\rm eV})$ in the atomistic picture.
In a recent experiment 
with macroscale samples \cite{CRAMER},
the stress is $\sigma \approx 10^1$ MPa and 
Eq.~(\ref{GRIFFITH-CG}) gives a macroscale length
($c_G \approx 1$ mm).
Since the length $c_{\rm G}$ 
is not dependent on the sample size $L$,
the fracture behavior can be expected to be different 
from the above picture,
in case that 
the sample size $L$ is smaller than  
the critical length $c_{\rm G}$ $(L < c_{\rm G})$.
In this letter, we will see such a situation 
in {\it nanocrystalline} silicon, 
in which the numbers of atomic layers
for these lengths 
$(\approx c_{\rm G}/d_0,L/d_0)$
are not macroscale numbers.

\begin{figure}[bt]
\begin{center}
   \includegraphics[width=5cm]{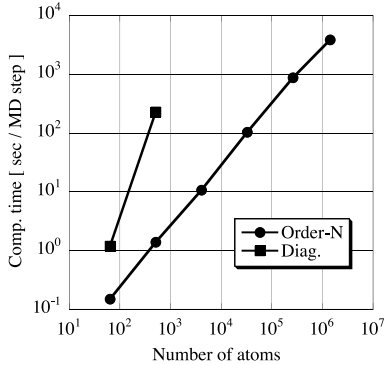}
\end{center}
\caption{
The computational time 
for bulk silicon as the function 
of the number of atoms, up to 1,423,909 atoms;
The CPU time is measured 
for one time step in the molecular dynamics (MD) simulation. 
A tight-binding Hamiltonian is solved using 
the perturbative order-N method and 
the exact diagonalization.
We use a standard workstation with 
single CPU and 2 GB RAMs.
}
\label{FIG-CPU-ORDER-N}
\end{figure}

For atomistic fracture simulations of silicon crystals,
a recent work of classical modelings \cite{HOLLAND-MARDER}
was done with $10^5$ atoms.
A more recent work\cite{GUMBSCH2}, 
however, pointed out 
the limited applicability of classical modelings
and the importance of electronic structure calculations. 
On the other hand,
there are several 
{\it ab initio} calculations with $10^2$ atoms
\cite{SPENCE93,GUMBSCH2}.
Due to the system size of simulations, 
these investigations are limited in situations, 
such as the preparation of 
the initial cleavage plane 
in which the reconstructed surface structure is assumed.
Therefore, 
large-scale electronic structure calculations
are essential.

So far, we have developed
several order-N methods for 
large-scale electronic structure calculations
\cite{HOSHI2000A}.
The order-N method is the general name of methods
in which the computational cost is proportional to the system size ($N$).
We have developed the variational and perturbative order-N methods
based on generalized Wannier states
\cite{HOSHI2000A,HOSHI2001A}.
The Wannier states $\{ \phi_i \}$ are localized
and the index $i$ denotes its localization center.
The equation for wave functions is common between the two methods
and is given by the one-body density matrix 
$(\rho \equiv \sum_i^{\rm occ.} | \phi_i \rangle \langle \phi_i |)$
\cite{HOSHI2000A}.
In the computational algorithm,
the perturbative method is simpler than the variational method.
Figure \ref{FIG-CPU-ORDER-N} demonstrates our large-scale calculation
among $10^2- 10^6$ atoms.

With the above two methods,  we now construct
a novel hybrid scheme, in which
the variational method is used only for  
the wave functions whose centers locate 
near fracture regions.
The regions contain, typically,
$4 \times 10^4$ electrons.
Some of such wave functions change 
their character dynamically from the bulk (sp$^3$ bonding) states
to surface ones, as discussed later.
The other wave functions, in bulk regions,
keep the character of the bulk bonding states and 
can be obtained by the perturbative method.
The wave functions $\{\phi_i \}$ calculated by the two methods 
are commonly used for constructing the one-body density matrix $\rho$.
Any physical quantity is expressed 
by the density matrix \cite{HOSHI2000A} and
is well defined in the present hybrid scheme.

The present work is based on 
a transferable tight-binding Hamiltonian
with s and p orbitals \cite{KWON}.
It is used for several crystalline phases 
and non-crystalline phases, 
such as liquid \cite{KWON} and surfaces \cite{SURFACE-TBMD}.
Since the fracture is 
the formation of surfaces in a bulk region,
the theory should reproduce 
the atomic structures 
both in bulk and surface phases,
which is satisfied in the present Hamiltonian.
The essence of the quantum mechanical freedoms
is the fact that 
the sp$^3$-hybridized bonds are formed in 
the bulk region,
while are not on surfaces.
To analyze the hybridization freedom,
a parameter $f_{\rm s}^{(j)}$ is defined,
for a wave function $\phi_j$, as
\begin{eqnarray}
 f_{\rm s}^{(j)} \equiv   
 \sum_{I} | \langle \phi_j | I s \rangle |^2
 \label{FSDEF},
\end{eqnarray}
where $| I s \rangle $ is the s orbital at the $I$-th atom.
For example, $f_{\rm s} = 1/4$ 
in an ideal sp$^3$ hybridized state.

In this letter,
we focus on the Si(001) surface, 
a standard template of the modern silicon technology.
A characteristic feature in the Si(001) surface 
is the formation of asymmetric dimers
\cite{CHADI79-SI001,RAMSTAD}.
The asymmetric dimer is connected by
a \lq $\sigma$' bonding state.
Another state is localized on the \lq up' atom, 
the dimerized atom near the vacuum region.
This localized state is called \lq $\pi$' state, 
because the direction of its p components 
is nearly perpendicular to the dimer bond.
Here an energy quantity is defined as
\begin{eqnarray}
 \Delta \varepsilon_{i}^{\rm (cov)}
 \equiv \langle \phi_i | H | \phi_i \rangle  - 
 \left[ f_{\rm s}^{(i)}  \varepsilon_{\rm s} + (1-f_{\rm s}^{(i)}) \varepsilon_{\rm p} \right]
 \label{COHE-ENE}.
\end{eqnarray}
A negative value of $\Delta \varepsilon_{i}^{\rm (cov)} $ corresponds to
the energy gain of a covalent bonding.
The \lq $\sigma$' state has the gain of 
$\Delta \varepsilon_{i}^{\rm (cov)}  \approx -2{\rm eV}$,
which mainly contributes to 
the dimerization energy (about $- 2{\rm eV}$)\cite{RAMSTAD}.
The \lq $\pi$' state 
has much smaller $\Delta \varepsilon_{i}^{\rm (cov)}$,
which is comparable to the energy difference 
between the asymmetric and  symmetric dimers
(the order of $0.1{\rm eV}$)\cite{RAMSTAD}.

\begin{figure}[b]
\begin{center}
  \includegraphics[width=7cm,clip]{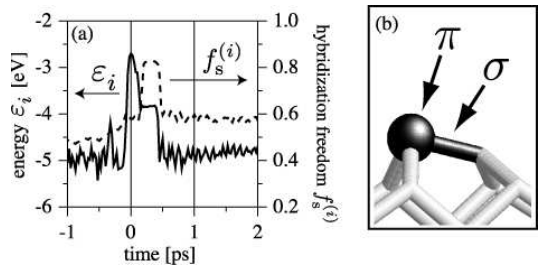}
\end{center}
\caption{
(a) Elementary reconstruction process 
with the one-electron energy 
$ \varepsilon_i $ and the weight on s orbitals $f_{\rm s}^{(i)}$. 
(b) An asymmetric dimer on a resultant crack.
The black rod and black ball correspond to
the \lq $\sigma$' and \lq $\pi$' states, respectively.
}
\label{FIG-E-PROCESS}
\end{figure}


\begin{figure*}[hbt]
\begin{center}
   \includegraphics[width=14cm]{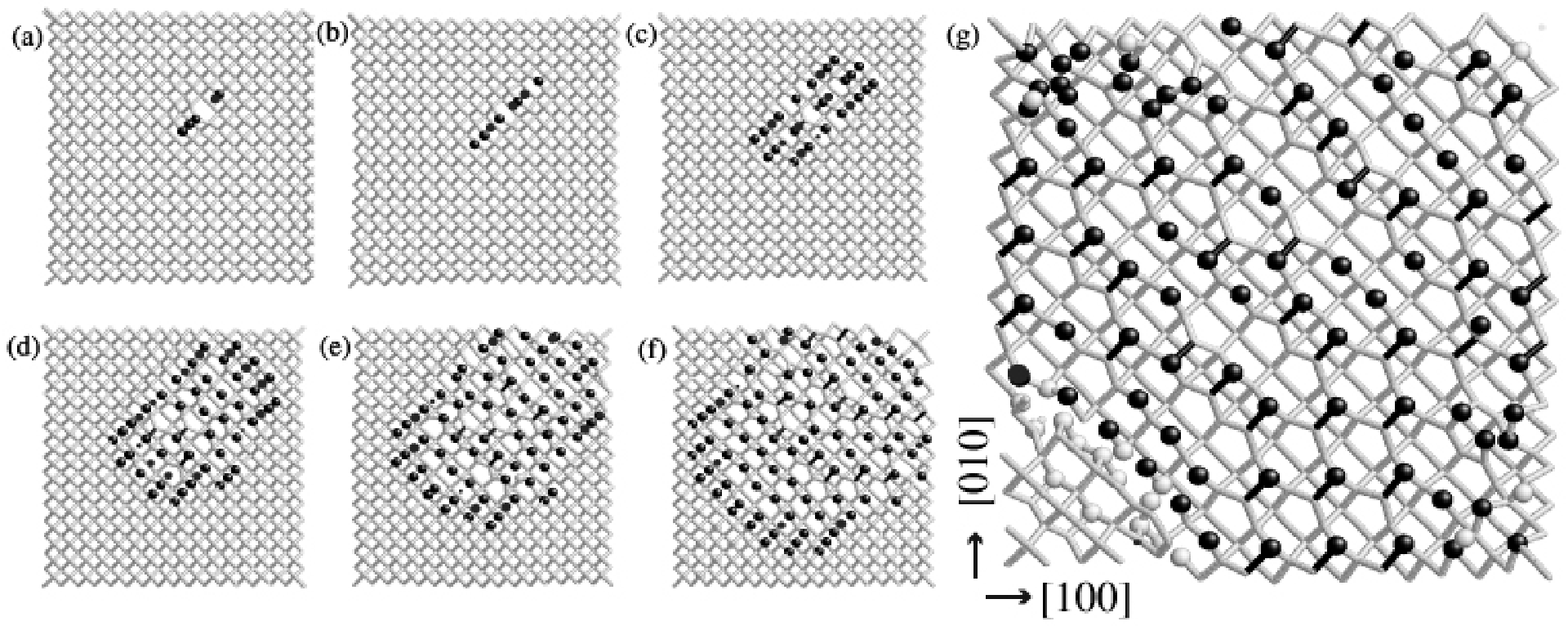}
\end{center}
\caption{
Snapshots of a fracture process in the (001) plane.
The sample size is 
$n_{100} \! \times \! n_{010} \! \times \! n_{001} \! = 
\! 33 \! \times \! 33 \! \times 33$ 
(4501 atoms).
The time interval between 
two successive snapshots is 0.3 ps, 
except that between (f) and (g) (about 1.3 ps).
A set of connected black rod and black ball
corresponds to an asymmetric dimer,
as in Fig.~\ref{FIG-E-PROCESS}(b) .
The left-down area has not yet fractured.
}
\label{FIG-ANIME}
\end{figure*}


The simulation details are as follows;
The hybrid order-N scheme is used for 
systems with 10$^4$ atoms or more.
In systems smaller than  the above size, 
the variational method is used in the whole region. 
The samples are isolated tetragonal clusters,
whose geometries are labeled with 
the number of atomic layers,
such as 
$n_{100} \! \times \! n_{010} \! \times \! n_{001}$ or 
$n_{110} \! \times \! n_{1\bar{1}0} \! \times \! n_{001}$.
As the boundary condition,
the Wannier states at the sample surfaces
are terminated by fixed sp$^3$ bonding states
and are not reconstructed.
The time step of the molecular dynamics is 3 fs. 
The total kinetic energy 
is controlled to be that with 300 K
by the Nose thermostat method \cite{NOSE}.
The numerical accuracy is checked 
among bulk, surface and fracture properties,
such as the elastic constants, 
the dimer formations on the clean (001) surface,
and the critical stress for fracture.
The last quantity is checked, with smaller samples,
in comparison with the standard diagonalization.
The calculated fracture propagation velocity 
is always in the same order of, but less than, 
the Rayleigh surface wave velocity ($4.5$km/s),
as expected from the continuum theory \cite{BRITTLE-TEXTBOOK}.

For fracture propagations,
external loads in the [001] direction are imposed.
During the simulations,
the external loads can be dynamically 
controlled by 
the atoms on the sample surfaces in the $z$ direction.
These atoms are fixed or 
under artificial constant-velocity motions in the $z$ axis.
The velocity,
typically $10^{-2}$ km/s, 
is much smaller than that of 
observed fracture propagation velocities 
(km/s).
As a seed of fractures,
a short range repulsive potential is imposed on 
one particular pair of atoms,
as a defect bond.
For smaller samples, 
the simulations begin 
without initial deformations.
The fracture always occurs
with the external loads 
in the order of $\sigma \approx 1$ GPa,
which corresponds to the strain energy of
$\sigma d_0^3 \approx 0.1{\rm eV}$ per atom.
For larger samples,
the simulations begin 
with initial static deformations 
in the above magnitude of external loads.  
The length $c_{\rm G}$ in Eq.~(\ref{GRIFFITH-CG})
is calculated as 
$c_G \approx 100 {\rm nm}$,
which is longer than 
the present sample sizes ($L \le 20{\rm nm}$).

In results,
a two-stage reconstruction process 
is commonly observed 
as the elementary process during 
successive bond breakings;
In Fig.~\ref{FIG-E-PROCESS}(a),
we monitor 
the one-electron energy
$\varepsilon_i \equiv \langle \phi_i | H | \phi_i \rangle $
and the hybridization freedom $f_{\rm s}^{(i)}$
of a Wannier state $ | \phi_i \rangle $.
Before the bond breaking 
($t \! <  \! 0 \, {\rm ps}$),
the wave function $| \phi_i \rangle $ 
is a bonding state in the bulk region, 
deformed due to the external load. 
At $t \approx 0 \, {\rm ps}$, 
a bond breaking occurs and 
the wave function  $| \phi_i \rangle $
loses the bonding character
with rapid increase of the bond length.
Then ($ 0 \, {\rm ps} \! < \!  t \! < \! 0.2 \, {\rm ps}$),
a twofold coordinated surface atom appears,
since another bond is broken almost simultaneously.
The wave function $| \phi_i \rangle $ 
forms a loan pair state 
that is stabilized by 
the increase of $f_{\rm s}^{(i)}$ $(0.6 \! \rightarrow \! 0.8)$.
The corresponding energy gain can be
estimated to be $-0.2 \! \times \!
( \varepsilon_{\rm p} \! - \! \varepsilon_{\rm s}) \approx -1.3$eV,
which explains the energy gain in the figure 
($\varepsilon_i \! = \!  -2.7 {\rm eV} \rightarrow  -3.8 {\rm eV} $).
In other words,
the bond breaking process is caused 
by the local electronic instability, 
that is, 
the energy competition between 
the {\it loss} of the bonding (transfer) energy and  
the {\it gain} due to 
the increase of the weight on the s orbitals ($f_{\rm s}$).
Finally, after the thermal motions with a finite time
 ($t \approx 0.4 {\rm ps}$),
a pair of  twofold coordinated atoms forms 
an asymmetric dimer with 
a $\sigma$ bonding state $| \phi_i \rangle$.
The corresponding covalent-bonding energy, 
defined in Eq.~(\ref{COHE-ENE}),
is $\Delta \varepsilon_{i}^{\rm (cov)}  \approx -1.9$ eV.
This energy explains the gain in the figure 
($\varepsilon_i \! = \! -3.8 {\rm eV} \rightarrow  -4.8 {\rm eV} $)
and the energy {\it loss} (about 1.3eV) 
due to the decrease of  $f_{\rm s}^{(i)}$
$(0.8 \rightarrow 0.6)$.
This asymmetric dimer is preserved 
until the end of the simulation, during a couple of pico seconds.
Figure ~\ref{FIG-E-PROCESS}(b) 
is an example of observed asymmetric dimers.

\begin{figure*}[hbt]
\begin{center}
 \includegraphics[width=14cm]{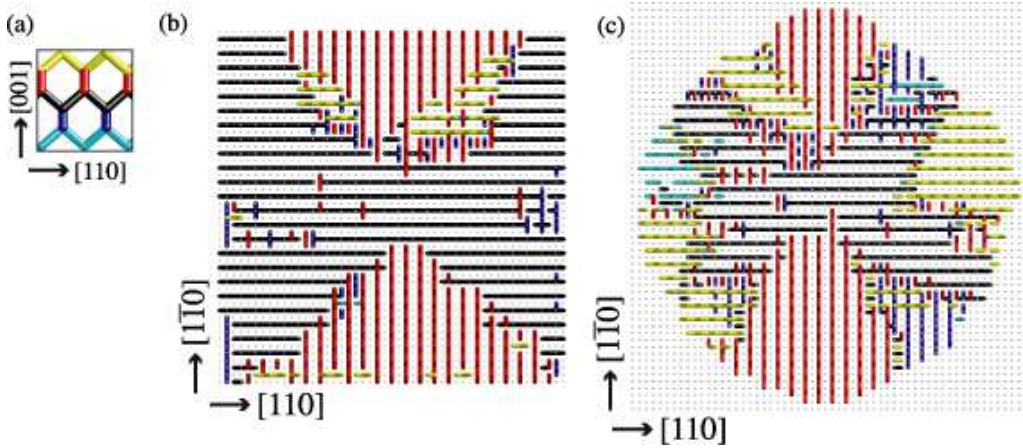}
\end{center}
\caption{
(a) Ideal diamond structure with colored bond sites.
(b)(c) Geometry of resultant cracks in the (001) plane. 
The broken bond sites are plotted as colored rods
in the ideal (crystalline) geometry. 
Rods within one layer 
are painted in a same color, as in (a).
The layer of black rods 
contains the defect bond at its central area.
Atoms are plotted as dots.
The sample sizes in (b) and (c) are
$n_{110} \! \times \! n_{1\bar{1}0} \! \times \! n_{001} \! 
= \!  49 \! \times \! 50 \! \times \! 49$
(30025 atoms) 
and $97 \! \times \! 100 \! \times \! 49$ (118850 atoms) , respectively.
In (c), only a central area 
($n_{110} \times n_{1\bar{1} 0} = 58 \times 60 $) 
of the sample is shown.
Note that 
the length of $n_{110}=50$ atomic layers
is about 10 nm.
}
\label{FIG-STEPS}
\end{figure*}


Figure~\ref{FIG-ANIME} shows the fracture process of  
a cubic sample with 4501 atoms.
Each Wannier state is classified
from its weight distribution
into a bonding or atomic orbital, 
which is shown as 
a rod or a ball in the figures, respectively. 
A {\it black} rod or ball corresponds to one in the layer 
that contains the defect bond.
One almost flat (001) surface is being created
with many asymmetric dimers.
The surface contains, however, 
many twofold coordinated atoms 
that have two back bonds (white rods)
and a lone pair orbital (black ball).
This is because 
a lone pair states
are metastable, as discussed above. 
In Fig.~\ref{FIG-ANIME}, 
an anisotropic bond-breaking propagation is seen in 
the [110] and [1$\bar{1}$0] directions,
especially in the early snapshots.
In the $[110]$ direction, 
the successive bond breakings propagate 
along the {\it nearest neighbor} bond sites, 
which forms a zigzag path, 
as the black rods in Fig.~\ref{FIG-STEPS}(a).
A bond breaking process
drastically weakens the {\it nearest neighbor} bonds,
due to the local electronic instability,
observed in Fig.~\ref{FIG-E-PROCESS}(a).
Therefore, the successive bond breakings 
propagate easily in the $[110]$ direction.
In the $[1\bar{1}0]$ direction,
on the other hand, 
the bond-breaking paths
are not connected,
as the red rods in Fig.~\ref{FIG-STEPS}(a).
In this direction,
the bond breakings are propagated 
through the local strain relaxation,
not by the local electronic instability.
As results,
the bond breaking propagation 
along the {\it nearest neighbor} bond sites
(in the $[110]$ direction of the present surface)
is faster than 
that in the perpendicular direction
(in the $[1\bar{1}0]$ direction),
due to the difference of the 
successive bond breaking mechanisms.
Note that a flat (001) surface 
is also obtained by a similar simulation 
{\it without the initial defect bond},
in which the fracture begins 
at the sample edges.

Figures \ref{FIG-STEPS} (b) and (c) show 
larger samples with step formations \cite{NOTE-STEP1}.
In the two cases,
all the conditions are the same, except the sample sizes.
To see the step structures clearly,
the broken bond sites are shown as rods 
in the {\it ideal} crystalline geometry.
The defect bond 
is located in the center of the drawn area. 
The anisotropic fracture propagation 
in one (001) plane
increases the anisotropic strain energy \cite{NOTE-ELAS-ISO}.
The anisotropy originates from 
the inequivalence 
between the [110] and [1$\bar{1}$0] 
directions within {\it one} (001) layer.
Since the above inequivalence 
does not appear within {\it two} successive layers,
a step formation between them
will release the anisotropic strain energy.
In Fig.~\ref{FIG-STEPS} (b), a step is formed 
between the layer of black rods and that of red rods.
In the $[110]$ direction, the bond-breaking 
propagation reaches the sample surfaces without step formations.
In the [1$\bar{1}$0] directions, 
the bond-breakings propagate slower and 
a step is formed in the central area  
at an early period of the crack propagation.
After that, 
the fracture propagates among the two atomic layers.
Since the two layers are symmetrically equivalent,
the resultant step formation path,
is almost a line in the [100] or [010] directions,
as the boundary of the fractured areas between the two layers.

In Fig.\ref{FIG-STEPS}(c),
the largest sample in the present letter,
the above line structure does not 
reach the sample surfaces
but is canceled
with additional step formations in complicated paths.
The sample size dependence of the step structures
is understood by 
the beginning of the crossover 
between nanoscale and macroscale samples; 
If the sample contains so many atoms,
the geometry of the resultant crack 
will be almost circular, 
as in Fig.~\ref{FIG-STEPS} (c),
so as to minimize the anisotropic strain energy 
\cite{NOTE-ELAS-ISO}.
If not, 
the strain energy is accumulated only within 
the confined bulk region due to the finite sample size.
The resultant fracture behavior is directly related
to the anisotropic atomic structure of the cleaved surface, 
as in Fig.~\ref{FIG-STEPS} (b).

Since the above mechanism of step formations
is two-dimensional,
the present samples may be
nanoscale \lq thin' samples.
In larger or thicker samples,
an expected fracture behavior 
is the bending of the fracture plane 
into the (111) plane,
the easiest cleavage plane in macroscale samples,
which is the crossover in the present context.
In an {\it enough} large sample,
the fracture mode with the easiest cleavage plane
will grow with no regard for sample shapes 
and details of conditions.
Note that the dynamical simulation with 10$^5$ atoms
is the practical limitation within a single CPU workstation 
and the program code with parallel computations
is now being developed for simulations with larger samples.

This letter shows 
a possible difference in fracture behaviors 
among nanoscale and macroscale silicon crystals.
Its origin is 
the size dependence of 
the energy competition 
between bulk and surface regions.
The electronic structures between the two regions
are essentially different 
and can be described  
by the present method with the well-defined total energy.
This energy competition is also inherent
in other phenomena,
such as crystal growth and self organizations, 
which may be candidates for applications
of the present method.


This work is supported by 
a Grant-in-Aid for COE Research \lq Spin-Charge-Photon' and
a Grant-in-Aid from the Japan Ministry
of Education, Science, Sports and Culture.
This work is also supported by
\lq Research and Development for 
Applying Advanced Computational Science 
and Technology' of Japan Science and Technology Corporation.



\begin{thebibliography}{99}
\bibitem{NOTE-ELAS-ISO}
The elastic property of silicon crystal shows 
only a small anisotropy within (001) plane;
the [$110$] and [$\bar{1}10$] directions are equivalent 
and the values of the Young modulus 
are different by only about 30 \% in 
the [100] and [110] directions.
\end{thebibliography}
\end{document}